\documentclass{article}
\usepackage[utf8]{inputenc}
\usepackage{amsmath}
\usepackage{amssymb}
\usepackage{amsthm}
\usepackage{url}

\title{Dynamics of Bitcoin mining}
\author{Nemo Semret\\
nemo@semret.org}
\date{January 15, 2022}
\begin{document}
\maketitle
\begin{abstract}
What happens to mining when the Bitcoin price changes, when there are mining supply shocks, the price of energy changes, or hardware technology evolves? We give precise answers based on the technical forces  and incentives in the system. We then build on these dynamics to consider value: what is the cost and purpose of mining, and is it worth it? Does it use too much energy, is it bad for the environment? Finally we extend our analysis to the long term: is mining economically feasible forever? What will the global hash rate be in 40 years? How is  mining impacted by the limits of computation and energy?  Is it physically sustainable in the long run? 
From first principles, we derive a fundamental scale-invariant feasibility constraint, which enables us to analyze the interlocking dynamics, find key invariants, and answer these questions mathematically. 
\end{abstract}

\section{Introduction: Proof-of-work}

"Proof-of-work" is the mechanism by which the Bitcoin ledger (known as the "blockchain") is maintained\cite{bitcoin}. Participants, called "miners", perform computations consisting of SHA-256 hash function evaluations. Whenever a miner finds an output value with a particular property (a certain number of leading zeros), that miner can create a candidate block to add to the chain, containing a batch of Bitcoin transactions. The updated chain is broadcast to the network, and the longest chain is accepted as the canonical chain by "consensus". When a block is accepted, the miner who created it earns a reward consisting of new BTC plus all the transaction fees for the transactions in the block. The hash function is such that the only way to find a valid output is brute force, trying different inputs randomly. The chance of success is simply proportional to the hash rate as a fraction of the global hash rate. There are no economies of scale in the mining function, if a miner has 10 times the hash rate of another, it will simply earn 10 times more on average. The number of leading zeros required in the hash output determines how rare the valid outputs are, and this "difficulty" is adjusted every two weeks so that a new block is created on average every 10 minutes.  To smooth out the randomness, miners typically work in "mining pools" where they share rewards proportionally to their hash rate. 

\section{Definitions}

Let \(\tau\) be the average time between blocks, \( N \) the number of miners, \(H \) be the global hash rate (hashes per second) and \( E \) the total energy used per block across all miners (following  electricity market convention, we use kWh as the unit of energy rather than Joules).  Mining revenue consists of two parts: an amount \(R \) in BTC automatically issued to whoever creates a block (currently \(R = 6.25\)), and  \(F\) in BTC, representing the total transaction fees in a block. Let \( p_b \) be the price of Bitcoin in \( \$ / BTC\).

The cost of mining consists of two main components, hardware and energy. For energy, we simply need the price of electricity \(p_e\) (\$/kWh). For hardware, the cost of mining machines depends on how powerful they are, how efficient they are and how long they last.   Since a machine has a finite life, we represent the cost as \emph{depreciation} -- the cost of a machine divided by its lifetime to get a value \(C \) in \(\$/s\). Note that the precise definition of a "machine", whether it is a single ASIC mining rig or many, is irrelevant, as long as we use the same definition as in \(N\) the number of miners. 

\section{Scale-invariant feasibility}

Mining is economically feasible if and only if revenue is not less than cost:
\begin{equation}
\label{revenue}
     p_b (R+F) \geq p_e E + N C \tau   
\end{equation}
or, equivalently,  
\begin{equation}
\label{feasible}
V_h \geq p_e  \alpha  + p_h 
\end{equation}
where 
\begin{itemize}
\item \( \alpha  = E / H / \tau \) is the hardware's efficiency in terms of energy use per computation in \( kWh/(hash/s)/s \) which is just \(kWh/hash\),
\item \(p_h = C/(H/N) \), depreciation divided by hash rate, is the hardware cost of computation in \( \$/s/(hash/s) \) which is  just \( \$/hash \),
\item \(V_h = p_b (R+F)/(H\tau) \) is the market's valuation of computation in \( \$/hash \).
\end{itemize}
On the right hand side, we have the variables that individual miners control, namely their cost of electricity and hardware.   While on the left hand side, we have the exogenous variables which together determine the market value of computation \(V_h\):  
\begin{itemize}
\item \(p_b\), the value attributed by users to the entire ecosystem,
\item \(F\), the demand for space on the blockchain, 
\item  \( H \), the competitive supply of mining capacity.
\end{itemize}

\(\tau \) and \(R \) are predetermined by the protocol.  The periodic difficulty adjustment acts as a feedback control loop which maintains \(\tau \) constant in the long term as \( H\) changes.  The feasibility equation (\ref{feasible}) is in the dimension of price of computation (\( \$ / hash \)). It is scale-invariant and applies to individual miners as well as globally.

\section{Dynamics}

The feasibility equation represents the boundary of a multi-dimensional dynamical system. Competitive mining creates forces  which push whichever variables have slack until  the system reaches this boundary. When conditions change the system moves to a new equilibrium point on the boundary. In other words the inequality always approaches equality,
\begin{equation}
\label{compet}
V_h = p_e  \alpha  + p_h  + \epsilon 
\end{equation}
where \(\epsilon\) represents the mining gross profit margin.
This occurs through a number of processes at different time scales.

\subsection{Bitcoin market}
\label{btc_market}
If the market price \(p_b\) goes up,  with no change in hardware technology, \(V_h\) and profit margins temporarily increase and more miners are incentivized to join, causing  \(H\) to rise, which in turn brings \(V_h \) and profit margins back down, resulting in a new equilibrium with higher \(H\). Thus, the amount of computation used to secure the blockchain grows proportionally to  the value of the network, but \(V_h\) remains stable.

\subsection{Supply shocks}
If  the global mining capacity \(H\) suddenly drops, e.g. because of a political change or a natural disaster affecting some miners, then the block time  \(\tau\) will immediately increase, since a lower hash rate means it takes longer to find new valid outputs, and \(V_h\) will initially remain the same. However, difficulty will gradually decrease until \(\tau\) is back to around 10 minutes, causing \(V_h\) to rise,  and the higher profit margins will draw in new miners that previously were kept out by their higher \(p_e\) or \(\alpha\), increasing \(H\) until a new  equilibrium is reached at a higher \(V_h\).

\subsection{Energy market}
\label{energy_market}
Miners that have a lower electricity price \(p_e \) will displace those for which it's higher.  The displacement occurs because profitable miners will add hash power, and as \(H \) increases, it reduces everyone's profit margins, and  less profitable miners will become unprofitable and drop out, resulting in a new equilibrium, with lower \(V_h\). A miner with more energy efficient hardware can still compete with a miner that that has cheaper electricity. But since there's no fundamental geographic constraint, over time, the population of miners migrates toward lower cost of electricity worldwide. 

\subsection{Hardware market}
Similarly, hardware that is  more energy efficient (lower \( \alpha \)) and cheaper (lower \(p_h \)) will displace less efficient and more expensive hardware.   The population evolves towards greater efficiency as microchip design and manufacturing progresses. Less efficient but cheaper hardware (higher \( \alpha \) and  lower \(p_h\)) can remain competitive: when a new generation with lower \(\alpha \)  becomes available, previous generation models drop in price; and existing miners for which the hardware cost is already sunk may continue to run as long as \(V_h \geq p_e  \alpha\).


\subsection{Current state}
\label{currentvalues}
Machines range from low-end machines (which we label OG) to high-end machines (NG). Current values\footnote{OGs are 10TH/s, 1kW machines, while NGs are 100TH/s, 3kW. We use a 4-year depreciation schedule. $p_h$ does not include  datacenter space, cooling, network access, and other operational costs.} rounded to one significant digit are:
\begin{center}
\begin{tabular}{ |l | l | l | l| }
\hline
       & OG & NG &  \\
\hline        
\(\alpha\) & 3 &  0.8 & \(10^{-5}\) ~kWh/TH  \\
\hline
\(p_h\)    & 0.6     & 1    & \(10^{-6} \) ~\$/TH \\
\hline
\(V_h\) & \multicolumn{2}{|c|} {3} & \(10^{-6}\) ~\$/TH \\
\hline
\end{tabular}
\end{center}
 As expected, the less efficient machines are also cheaper, as \(p_h\) of machines on the secondary market adjusts in response to \(\alpha\) of newer machines.  Even if the \$/BTC price changes rapidly, per the dynamics above, \(V_h \) should remain stable, evolving at the same pace as \(\alpha\).
 
\section{Value of mining}

\(V_h\) represents the market's valuation of computation, and therefore the upper bound on the cost of computations.   The market value  of  mining activity as a whole is \( V_h H\), which is a \$ value per unit of time. This value is ultimately paid by the users, directly through  transaction fees \(F\) plus indirectly via the inflation caused by new coins \(R\).  

Why do users give this value? Mining provides the ability to store and transfer value. But this ability doesn't depend on the specific quantity of computation \(H\) since, thanks to the difficulty adjustment, the network could handle just as many transactions and store just as much value at any level of \(H\).  In other words, while \(H\) is exogenous to a single miner (it represents competition from other miners), it is endogenous to the system as a whole (it's just a function of other forces in the system rather than a fundamental "input" from users). There's no inherent direct benefit of \(H\) but if a malicious entity controlled the majority of the hash rate (a "51\% attack"), it would be able to  "double-spend", making current transactions  untrustworthy. If the attacker can persist long enough, users will no longer trust they can transact in the future, and Bitcoin is ruined. The system  therefore needs \( V_h H\) to be large enough that no single miner has more than half of it. 

Thus the macro value of mining is best understood as the users continuously paying the miners a fraction of the value in the system, large enough to keep it decentralized (see section \ref{decentralization} for an elaboration).   This fraction is:
\begin{equation}
\label{value}
     V_t = { V_h H   \over p_b M} = {R+F \over M\tau}
\end{equation}
where \(M \) is the total amount of BTC in existence. 

Note that neither the energy efficiency (\(\alpha\)), nor the cost of hardware (\(p_h\)), nor the amount of computation (\(H\)), nor the price of energy (\(p_e \)) matter to \(V_t\). For an individual miner those variables are vital. But for the system as a whole, however those variables may change, the share of value captured by mining will remain near the same equilibrium. 

In other words, the macro cost of mining is not a matter of technology, it is purely a socioeconomic equilibrium. Currently,
\begin{equation}
\label{valuerel}
  V_t= 1.8 \%/year
\end{equation}
but it will change as \(R+F\) changes in the long run (more on that in section \ref{long_term_fees}). 

\section{Value judgement}

Now let us consider whether the current share of value captured by mining is sustainable.  Recall that it  has to be high enough to keep the system decentralized but low enough for users to be willing to pay it in exchange for the service of storing and transferring value.

\subsection{Decentralization as security}
\label{decentralization}
First consider whether \(V_t\) is currently high enough to provide the desired security. The key question is what is the size of a realistic threat? The cost needs to be large enough to be prohibitive for any single attacker (decentralization).

Consider the attacker's point of view, assuming the motivation is profit. First getting the majority of the hash rate, with current values as in section \ref{currentvalues}, translates to roughly \$25B of equipment up front (or more as hardware supply gets squeezed and raises $p_h$), and electricity at a rate of \$4B/year. Second, the attacker would have to successfully perform double-spending transactions worth a much greater amount to recoup their cost and make a profit. Third, they would have to successfully extract it by exchanging it for other currencies, goods or services. Finally, all of that would have to be executed rapidly before the value of BTC collapses as a result of the attack, and obviously the attack can only be done once. By contrast, simply continuing to mining honestly would generate \$7B per year, potentially for a long time. Thus honest participation is more profitable than stealing via a 51\% attack.

What if the attacker only wants to destroy the system rather than profiting from an attack directly?  Recall the attack has to be sustained, so it would face various defensive reactions. One possible defense is a "hard fork" that obsoletes the current hardware, and draws the market value out of the existing blockchain. With ordinary hard forks (like BCH and BSV for example), the existing blockchain of BTC tends to keep the market due to the overwhelming network effect. But in our attack scenario, there there would be a reason for users to try to move to a new chain. This defense would of course be very costly for the users, but the attacker would have to start from scratch several times to win.   A much simpler and more likely defense is other entities countering the initial attack with an increase in hash rate to defend the whole stored value and they would be willing to go much higher than 1.8\% per year, even at a temporary loss. Thus while the amount  may seem small for a global system (1.8\% is about \$14B/year currently), when looking at it game theoretically, it appears  an attack would have to sustain many times that rate to be successful.  

Overall, the current equilibrium  \(V_t\) seems to be sufficient to deter "51\% attacks", and empirically that has been the case for the last 13 years. Further, while at the current size, the largest realistic threat is proportional to the target, it is ultimately capped by the largest possible offensive budget of a single entity. Thus if the market cap of Bitcoin grows much larger, the threat will have to decline as a percentage, and this aspect of security should improve.

\subsection{Cost to users}
\label{usercost}

Second, let's consider whether \(V_t\)  is too much. Of course, if one believes that Bitcoin itself has no value, then anything greater than zero is too much. Conversely, the people who use Bitcoin must believe it is worthwhile since they are paying it.  But this belief is largely implicit and highly speculative, so it's natural to ask whether 1.8\%/year is a reasonable share for the service provided by miners.  

For reference, one can do "thought experiments" comparing it to similar costs as a fraction of total wealth in various contexts involving say, storing and transmitting gold or cash. For example, gold storage vaults typically charge, per year, around 0.5\%-1\% of the value stored, and sending gold safely around the world safely requires expensive manpower.  For cash, international money transfer services often cost around \$20 per transaction, and for many recipients of remittances, these fees add up to more than 2\% of their wealth per year.  Comparisons with credit cards, Paypal and the like, which cost about 2-3\% per transaction, are more complicated as these services depend on accounts with an intermediary institution. This is akin in the Bitcoin world to "custodial" Bitcoin services (companies like Coinbase, or Square), or better yet with the Lightning Network\cite{lightning} which decentralizes the intermediary function as a "layer 2" service built on top of the Bitcoin blockchain as the base layer. In these cases, transactions are faster and lower-cost indeed, but there's a trade-off with full settlement and custody being deferred.    Overall, despite the early stage uncertainty, the current cost to users doesn't seem irrational when compared to other systems.

\subsection{Does mining use too much energy?}

Leaving aside the cost to users, a more frequently raised question is whether Bitcoin uses too much or wastes energy, in a broader societal sense.  

This question assumes the system requires some amount of computation to be done and that it "wants" to minimize the energy to achieve it. That is indeed how most systems work.  But not Bitcoin. Proof-of-work does the reverse. The system "wants" a certain value to be spent on energy, and the amount of computation adjusts to achieve it. Of course individual miners compete by being as efficient as possible, but the resulting collective behavior is to achieve a certain cost of energy with variable amounts of computation, not to perform a certain amount of computation with variable amounts of energy. This unusual combination -- individual participants being  efficiency-seeking but their collective behavior being efficiency-neutral -- is very counter-intuitive and probably the root cause of much misguided hostility. 

It's also worth emphasizing that the amount of energy doesn't matter, only the cost. If the price of electricity relative to everything else in the world doubles, but nothing else changes, then Bitcoin would simply use half the amount of energy to achieve the same relative cost \(V_t\). Maintaining a certain relative  cost of energy is a feature not a bug, and "waste"  is impossible by design. All of the energy is "work". 

And where there's no "waste", the question of energy use boils down to a moral judgement.  Can you argue that heating in the winter, even if perfectly efficient, is not justified and people should move to warmer climates? What about air conditioning, electric clothes dryers, or ice cream?  When is any purposeful energy use  justified?  Morally, as long as access to and the price of energy is objectively fair, what it's used for should be accepted as a subjective choice. Bitcoin offers inflation-resistant savings, low-cost long-distance value transfer, and  censorship-resistant money, benefits which for its users are (subjectively) worth the energy.

\subsection{Carbon emissions}

Besides the question of quantity, a separate value judgement can be made on the sources of electricity, since not all forms of electricity generation are the  same when considering  "externalities" like carbon emissions. 

Many sources of renewable energy are highly variable: solar and wind power depend on time of day and weather,  hydroelectric power is seasonal,  etc.  In general, these ups and downs on the supply side do not line up perfectly with the demand for electricity. Further, even with the largest possible batteries, water reservoirs, etc., electric energy remains extremely difficult to store for later use at a large scale. Thus there is often a lot of "stranded" energy when using renewable sources. Just like off-peak bandwidth in telecommunication networks, or empty seats on scheduled airline flights, the cost of production is already sunk, and so for the supplier, selling stranded power at any price\footnote{In fact, sometimes even at a negative price, as in some cases suppliers who can't turn down production pay to offload excess power that could damage infrastructure if unused.} is better than letting it go unused.  

As mentioned in section \ref{energy_market}, the competitive dynamics of Bitcoin mining are such that it shifts in time and space to the lowest available cost of electricity. This occurs not just by deploying hardware to various locations, but also by turning miners on or off instantly. This flexible demand-side support makes mining the ideal customer to balance variable supply, and as variability tends to affect renewable much more than fossil fuel sources, in effect,  Bitcoin  subsidizes the development of "green" electricity.

\section{Long term trends}

\subsection{Transaction fees}
\label{long_term_fees}
Currently, of the two components of miner revenue, block rewards \(R\) are much greater than transaction fees \(F\). The design explicitly relies on \(R\) to incentivize mining in the early stages. As \(R\) is automatically  halved every 4 years, such that the total supply converges to 21 million BTC, the assumption is that \(F\) will rise to cover mining\cite{bitcoin}. 

The expected evolution of \(F\) is best understood as a congestion pricing mechanism.  The block size and the time between blocks are (roughly) constant, so we have about 1MB/10min = 1.67 kB/s of bandwidth for transactions. A typical transaction is a few hundred bytes, giving a maximum throughput on the order of 10 transactions per second. Fees are a way to get priority in this limited pipe. Thus $F$ ultimately depends not on the absolute value of Bitcoin, but  on the velocity of money in the network. In principle, we could see high value and low velocity, low value and high velocity, both high or both low.  Extrapolating from the past is difficult at this point because the velocity in the early stages of an emergent economic system, when much of the activity is speculative, might be fundamentally different from that when it matures. Further, the launch in mid-2017 of SegWit\cite{segwit}, which significantly improves space efficiency in the blockchain, complicates any historical analysis of demand for transaction bandwidth.

Nonetheless, we can see that in the early days, transaction volume was low and blocks were rarely full, so fees were near zero, as expected. Since mid-2012, as the price of BTC in \$ has risen by four orders of magnitude, both transaction fees  and transaction values have gone up and down within one order of magnitude (respectively 10 to 100 BTC/day, and 50k to 500k BTC/day). 

If the system succeeds as a global store of value,  there will certainly be demand for more than 10 transactions per second. By choosing the ones with the highest fees to include in each block, the miners are effectively conducting an auction for transaction bandwidth. Transactions whose fees are consistently  below the clearing price will adapt to use layer 2 services like Lightning\cite{lightning}, and the overall flow of transactions will be partitioned between the base layer (i.e. directly on the blockchain) and layer 2. The boundary between the two is not preset, it is an adaptive market equilibrium. If adoption increases, the induced fees on the blockchain will be higher, and as discussed in section \ref{decentralization}, if the market cap increases the mining share of value needed to keep it decentralized will be lower. Thus adoption risk is the biggest factor, but if Bitcoin grows, the incentives are favorable to decentralization of mining.


\subsection{Computing efficiency}
Currently, the two components of the cost of computation   \(p_e \alpha\) and \(p_h\) are roughly of the same magnitude.  They have declined as mining has evolved from running on CPUs, to GPUs, FPGAs and now ASICs. Each type of microchip, and each generation within a type, is more efficient at performing hashes than its predecessors. The rate of progress in the semiconductor industry for the last half century has been such that transistor density on microchips doubles every 18 months, and performance/cost follows the same exponential curve (Moore's law).  Whether efficiency gains in hash computation will continue as this rate or slow down is debatable, but regardless of the rate,  \(\alpha\) and \(p_h\) will likely continue to decline. However they will never be zero. Even if, for the sake of argument, we assume hardware costs \(p_h \to 0\), the value of \(\alpha \) has a non-zero lower bound given by the laws of physics. Landauer's principle\cite{landauer} states that each bit of output must consume \(kTln(2)\) Joules, where \(k\) is the Boltzman constant and \(T\) is temperature in Kelvin. With 256 bits per hash value and room temperature, we get a limit of 
\begin{equation*}
  \alpha \approx  2 \times 10^{-13} kWh/TH.  
\end{equation*}
Comparing that to current values in section \ref{currentvalues} shows that mining can get at most 100 million times more efficient. If we assume Moore's law, the limit may be reached in about 40 years.  

To get an upper bound on $H$, assume the mining share of value stays at  \(V_t = 1.8\%/year\) as in (\ref{valuerel}), \(M \to 21 \) million,   \(p_h \to 0\),  and maximum competition results in \(\epsilon \to 0\), then from equations (\ref{compet}) and (\ref{value}) we get: 
\begin{equation}
\label{hashrate}
 H \to { V_tM \over \alpha}{p_b \over p_e}  \approx 6 \times 10^{10} {p_b \over p_e } \quad  TH/s.  
\end{equation}
\(p_e/p_b\) is simply the price of electricity in BTC, currently \(\approx 10^{-6} BTC/kWh\), which gives an upper bound of \(6 \times 10^{16}\) TH/s.

Note that the \$ dimension cancels out. So even though \(H\) grows, the naive argument that if Bitcoin succeeds, then it could end up using a huge fraction of the world's energy (an argument which typically includes dramatic comparisons  with the electricity usage  of entire countries) is incorrect. The correct intuition is that the global hash rate is ultimately proportional to the value of Bitcoin relative to electricity. In the maximal scenario, it will tend to the value of money (as a store of value and means of exchange) relative to the value of energy in general. 

\subsection{Unit of account}
Prices \( (p_h,  p_e, p_b)\) are in the generic unit \$, and we used USD wherever there are actual values. However, there's nothing canonical about the USD, it could be any currency. If BTC itself becomes the unit in which electricity and hardware are priced, then all of the dynamics are the same except that we happen to have \( p_b = 1 \).  

\section{Conclusion}

Bitcoin combines ideas from cryptography, game theory, control systems, distributed algorithms and physics into a well balanced dynamical system. This system enables a remarkable innovation: a database that no one owns but everyone can access, which establishes truth without a trusted intermediary, and serves as a store of value and means of exchange without a central authority. In other words, Internet-native money, backed by energy.

\end{document}